\documentclass[nofootinbib,twocolumn,aps,letterpaper,superscriptaddress,showpacs]{revtex4}
\usepackage{amsfonts}
\usepackage{amssymb}
\usepackage{amsmath}
\usepackage{indentfirst}
\usepackage[brazil]{babel}
\usepackage[latin1]{inputenc}

\begin{document}

\newcommand{\be}{\begin{equation}} \newcommand{\ee}{\end{equation}}
\newcommand{\bea}{\begin{eqnarray}}\newcommand{\eea}{\end{eqnarray}}

\renewcommand{\thesection}{\arabic{section}}
\renewcommand{\thesubsection}{\thesection.\arabic{subsection}}
\renewcommand{\theequation}{\arabic{equation}}
\newcommand{\pv}[1]{{-  \hspace {-4.0mm} #1}}

\def\A{\mathcal{A}}
\def\C{\mathbb{C}}
\def\N{\mathbb{N}}
\def\R{\mathbb{R}}
\def\Z{\mathbb{Z}}
\def\g{\mathfrak{g}}
\def\h{\mathfrak{h}}
\def\p{\partial}
\def\d{\delta}
\def\proof{\noindent \textbf{Proof. }}
\def\qed{$\blacksquare$}
\def\F{\mathcal{F}}
\def\G{\mathcal{G}}
\def\H{\mathcal{H}}
\def\K{\mathcal{K}}
\def\x{\hat x}

\def\BI{{\rm 1\!l}}

\title{
{\small\hfill SINP/TNP/2011/07 }\\
Quantum statistics and noncommutative black holes}
%\author{}

\author{Kumar S. Gupta \footnote {kumars.gupta@saha.ac.in}}

\affiliation{Theory Division, Saha Institute of Nuclear Physics, 1/AF
Bidhannagar, Calcutta 700064, India}

%\author{S. Meljanac \footnote {meljanac@irb.hr} and A. Samsarov
%\footnote{asamsarov@irb.hr}}

\author{S. Meljanac \footnote {meljanac@irb.hr}}

\author{A. Samsarov \footnote{asamsarov@irb.hr}}

\affiliation{Rudjer Bo\v{s}kovi\'c Institute, Bijeni\v cka  c.54, HR-10002
Zagreb, Croatia}

\vspace*{1cm}

%\maketitle
\begin{abstract}
We study the behaviour of a scalar field coupled to a noncommutative
black hole which is described by a $\kappa$-cylinder Hopf algebra. We
introduce a new class of realizations of this algebra which has a
smooth limit as the deformation parameter vanishes. The twisted flip
operator is independent of the choice of realization within this
class. We demonstrate that the $R$-matrix is quasi-triangular up to the first order in the deformation parameter. Our results indicate how a scalar field might behave in the vicinity of a black hole at the Planck scale.\\
 
\noindent
Keywords: $\kappa$ deformed space, noncommutative black holes, twisted statistics

\end{abstract} 

\pacs{ 11.10.Nx, 11.30.Cp}

%\keywords {$\kappa$ deformed space, noncommutative black holes, twisted statistics }

\maketitle

%\newpage
\section{ Introduction}
Noncommutative geometry offers a framework for describing the quantum structure of  space-time at the Planck scale \cite{majid}. Einstein's theory of general relativity together with the uncertainty principle of quantum mechanics leads to a class of models with space-time noncommutativity \cite{dop1,dop2}. The smooth space-time geometry of classical general relativity is thus replaced with a Hopf algebra at the Planck scale. There are many examples of such Hopf algebras including the Moyal plane, $\kappa$-space and Snyder space. The analysis of \cite{dop1,dop2} does not suggest any preferred choice among these models. 

Further insight about the possible features of the space-time algebra at the Planck scale comes from the analysis of noncommutative black holes. The algebraic structure associated with a noncommutative black hole can be revealed by studying a simple toy model, such as the noncommutative deformation of the BTZ black hole \cite{brian1,brian2}. The resulting space-time algebra resembles a noncommutative cylinder \cite{C1,C2}, belonging to the general class of $\kappa$-deformed space-time \cite{L1,L2,L3,L4}. The appearance of the $\kappa$-cylinder algebra is not restricted to the deformation of the BTZ black hole alone. Such an algebra describes noncommutative Kerr black holes \cite{schupp} within the framework of twisted gravity theories \cite{wess1,wess2,seckin2d}. It also appears in the context of noncommutative FRW cosmologies \cite{ohl}. In addition, the $\kappa$-Minkowski algebra is relevant in models of doubly-special relativity and in the analysis of astrophysical data
  from the GRB's \cite{majid1,dsr1,glik1,glik2,glik3,Ghosh:2006cb,Ghosh:2007ai,lee1,lee2,rim,AmelinoCamelia:2011bm,AmelinoCamelia:2011cv,Gubitosi:2011ej,us3}. This wide-ranging appearance of the $\kappa$-cylinder algebra suggests that it captures certain generic features of noncommutative gravity and black holes and is therefore an interesting toy model to explore Planck scale physics.

In this Letter we shall investigate certain features of the $\kappa$-cylinder algebra using a scalar field as a simple probe. In order to study quantum field theory in any space-time, it is essential to specify the statistics of the quantum field. It has been known for a long time that quantum gravity can admit exotic statistics \cite{sorkin,an1,an2}. More recently, the idea of twisted statistics and the associated $R$-matrices have appeared in the context of quantum field theories in noncommutative space-time \cite{S1,S2,S3,S4,S5,S6}, including the $\kappa$-deformed spaces
 \cite{KS1,Govindarajan:2008qa,Govindarajan:2009wt,KS2,KS3,KS4,KS5,luk1,luk2,luk3}.
 To this end, it is useful to work with realizations of the
 $\kappa$-space and the associated star products
 \cite{Meljanac-2,Meljanac-3,Meljanac-5,Meljanac:2010ps,Meljanac:2011ge}. In this
 letter we introduce a new class of realizations of the
 $\kappa$-cylinder algebra which has a smooth limit as the deformation
 parameter vanishes, which is different from the previous discussion
 in the literature \cite{C1,C2}. In addition, we obtain the twisted
 flip operator and show that it is independent of the particular
 choice of realization within the class of realizations considered
 here.  

For the commutative theories of gravity, probing a black hole space-time with a scalar field yields rich information about the underlying geometry \cite{ssen1,ssen2,ssen3}.  In our context, this amounts to studying a scalar field coupled to the $\kappa$-cylinder algebra \cite{KS3,luk1,luk2,luk3,Govindarajan:2009wt,dice10}. Here we analyze this problem for a class of realizations of the $\kappa$-cylinder algebra that has a smooth limit as the deformation parameter vanishes. We find that the creation and annihilation operators appearing in the mode expansion of the scalar field satisfy a twisted oscillator algebra. This provides an initial glimpse as to how a scalar field might behave in the vicinity of a black hole at the Planck scale. 

This Letter is organized as follows. In Section 2 we review certain
features of the noncommutative black holes that are relevant for our
work. In Section 3 we discuss the realizations of the
$\kappa$-Minkowski algebra and introduce a new class of realizations
for the $\kappa$-cylinder algebra that has a smooth limit as the
deformation parameter vanishes. We also discuss the star products and
the twists associated with this algebra. In Section 4 we discuss the
twisted statistics and the
 $R$-matrices for the realizations discussed here. The twisted oscillator algebra for the class of realizations considered in this Letter is derived in Section 5. In Section 6 we conclude the paper with some discussions and an outlook.

\section{Noncommutative black holes}

We start by reviewing certain features of noncommutative BTZ black hole \cite{brian1,brian2} which are useful for our analysis. In the commutative case, the metric for a non-extremal BTZ black hole in terms of the Schwarzschild like coordinates $(r, \phi, t)$ is given by \cite{btz1,btz2}
\bea 
ds^2 &=& \biggl( M - \frac {r^2}{\ell^2} - \frac{J^2}{4r^2}\biggr) dt^2 
  +  \biggl( -M + \frac {r^2}{\ell^2} +
\frac{J^2}{4r^2}\biggr)^{-1} dr^2  \nonumber \\
& +& r^2 \biggl(d\phi - \frac J{2r^2} dt\biggr)^2
\;,
\label{btzmtrc}
\eea
where $0\le r<\infty\;, \;-\infty < t<\infty\;, \;0\le \phi<2\pi\;,$ and $M$ and
 $J$ are respectively the mass and spin of the black hole and  $\Lambda= -1/\ell^2$ is the cosmological constant. Here we restrict our attention to the non-extremal case, where the two horizons $ r_{\pm}$ are given by
\be 
r_\pm^2 =\frac {M\ell^2}2 \biggl\{ 1\pm \bigg[ 1 - \biggl(\frac
J{M\ell}\biggr)^2\biggr]^{\frac 12} \biggr\}. \;
\label{rpm}
\ee

The BTZ black hole can also be obtained by quotienting the manifold
$Ads_3$ or $SL(2,{\mathbb{R}})$ by a discrete subgroup of its isometry. A noncommutative version of the BTZ black hole can be realized by a quantum deformation of $Ads_3$ or $SL(2,{\mathbb{R}})$ in such a way that is compatible with the quotienting \cite{brian1,brian2}. In the resulting noncommutative theory, the coordinates $r$, $\phi$ and $t$ are replaced by the corresponding operators $\hat r$, $\hat \phi$ and $\hat t$ respectively, which no longer commute but satisfy the algebra
\be 
[e^{i\hat \phi},\hat t] = \theta e^{i\hat \phi}\qquad [\hat r,\hat t] =  
[\hat r,e^{i\hat \phi}] =0\;,
\label{qntmalg} 
\ee
where the constant $\theta$ is proportional to $\ell^3/(r_+^2-r_-^2)$. The algebra (\ref{qntmalg}) is so constructed that its cetral elements are kept invariant under the action of the isometry group of the BTZ black hole \cite{brian1}. We therefore see that the noncommutative BTZ black hole is equivalent to a noncommutative cylinder. 

From (\ref{qntmalg}) we find that the operator $\hat r$ is in the centre of the algebra.
What is perhaps not so obvious is that the operator $e^{-2\pi i\hat t/\theta}$ also belongs to the centre of this algebra. As a result, in any irreducible representation of (\ref{qntmalg}), the element $e^{-2\pi i\hat t/\theta}$ is proportional to the identity,
\be
e^{-2\pi i\hat t/\theta} = e^{i\sigma}\BI,
\label{central}
\ee
where the constant parameter $\sigma \in {\mathbb{R}}$ mod $(2 \pi)$. Eqn. (\ref{central}) implies that in any irreducible representation of (\ref{qntmalg}), the spectrum of the $\hat t$ operator is quantized \cite{C1,C2,paulo} and is given by
\be 
{\rm {spec}}~ {\hat t} = n \theta
-\frac {\sigma\theta}{2\pi}\;,\;\;n\in
{\mathbb{Z}}.
\label{dsctsptmt}
\ee

As mentioned before, a similar algebra as in (\ref{qntmalg}) appears in the noncommutative generalization of the rotating Kerr black hole \cite{schupp} as well as for the noncommutative version of the FRW cosmology \cite{ohl}. For the rest of this paper we shall consider the algebra in (\ref{qntmalg}) as a prototype for  noncommutative black holes and study its properties.

\section{Realizations of $\kappa$-Minkowski space}

\subsection{Generalities}

We start by recalling certain general properties of the $\kappa$-Minkowski algebra relevant for our analysis. The operators corresponding to coordinates on $\kappa$-deformed noncommutative space satisfy the relation
\begin{equation} \label{1}
  [\hat x_{\mu}, \hat x_{\nu}]  =  i(a_{\mu} \hat x_{\nu} - a_{\nu} \hat x_{\mu}),  
\end{equation}
where $ a_{0},a_{1},a_{2},...,a_{n-1}$ appearing in (\ref{1}) are  real constant parameters describing a deformation of Minkowski space. It is understood that the Greek indices run through the set $\{0,1,\ldots ,n-1\}$, and the latin indices run through the subset $\{1,2,\ldots ,n-1\}$ with the summation over repeating indices assumed. 

Next we introduce derivative operators with the properties 
\begin{eqnarray} \label{2}
&  & [{\p}_{\mu}, {\p}_{\nu}]   =   0, \\
 &  & [{\p}_{\mu}, \hat{x}_{\nu}]   =   {\phi}_{\mu \nu}(\p),
\end{eqnarray}
where the functions $ {\phi}_{\mu \nu}(\p)$ in derivatives here provide
realizations of noncommutative coordinates in the auxiliary Hilbert
space ${\mathcal{H}} ({\mathbb{R}}^{1,3}, d^4 x)$,
 which is spanned by the ordinary commutative coordinates 
\begin{equation} \label{4}
 [x_{\mu}, x_{\nu}] = 0, \qquad   [{\partial}_{\mu}, x_{\nu}] = {\delta}_{\mu \nu}.
\end{equation}
We also define a shift operator which satisfy
\begin{equation} \label{3}
 [Z, \hat x_{\mu}] = ia_{\mu} Z,   \qquad   [Z, \p_{\mu}] = 0.
\end{equation}
The realizations of noncommutative coordinates can then be written in the form \cite{Meljanac-2,Meljanac-3}
\begin{equation} \label{5}
 \hat{x}_{\mu} = x^{\alpha} {\phi}_{\alpha \mu} (\partial).
\end{equation}
We restrict our attention to deformation parameters with zero spatial components, $a_\mu = a_0 \delta_{0\mu}$, $a_0\neq 0$.
Then the commutation relations become
\begin{equation} \label{28}
[\hat x_i,\hat x_j] = 0,  \qquad [\hat x_0,\hat x_j]  = ia_0\hat x_j,  
\end{equation}
 The algebra (\ref{28}) admits the realization of the type (\ref{5}),
 which also includes the class of noncovariant realizations given by  \cite{Meljanac:2010qp}
\begin{align}
\hat x_0 &= x_0 \psi(A)+ia_0  x_k \p_k \gamma(A), \label{29} \\
\hat x_i &= x_i \varphi (A), \quad  \gamma=\frac{\varphi^\prime}{\varphi}\psi+1, \label{30}
\end{align}
where $A=-ia_0\p_0$.
 The above realization is parametrized by two real-analytic functions $\varphi$ and $\psi$ satisfying
the boundary conditions $\varphi(0)=\psi(0)=1$ and $\varphi^\prime
(0)$ is finite. The shift operator has the form
\begin{equation}\label{34}
Z=e^{\Psi(A)}, \quad \Psi(A)=\int_0^A \frac{dt}{\psi(t)}.
\end{equation}

%%%%%%%%%%%%%%%%%%%%%%%%%%%%%%%%%%%

The derivatives $\p_\mu$ generate the action on $\kappa$-space coordinates, which, together with angular momenta form a deformed Poincar\'{e} algebra that has a Hopf algebra structure. The deformed coproducts are given by
\begin{equation}
\Delta A = \Psi^{-1}\circ \ln (Z\otimes Z).
\end{equation}
and
\begin{equation}\label{63}
\Delta \p_0 = \frac{i}{a_0} \Delta A, \quad
\Delta \p_i = \varphi(\Delta A) \left(\frac{\p_i}{\varphi(A)}\otimes 1
 + Z\otimes \frac{\p_i}{\varphi (A)}\right), 
\end{equation}
\begin{equation}
 \Delta Z = Z\otimes Z. \nonumber
\end{equation}
We note that the coproduct of $M_{ij}$ is undeformed.
%The coproduct of the shift operator is simply , and the antipode is found to be
%\begin{equation}
%S(\p_0) = -\p_0, \quad   S(\p_i) = -Z^{-1}\p_i.
%\end{equation}
%The antipode of the shift operator is $S(Z)=Z^{-1}$. 
% Also, the counit for
%all the generators is undeformed. 

%%%%%%%%%%%%%%%%%%%%%%%%

\subsection{Smooth realizations of the $\kappa$-cylinder algebra}

We now focus on the algebra relevant for the noncommutative black hole introduced in Section 2.  Our aim is to investigate a version of $\kappa$-deformed
noncommutative space, where the only space coordinate is compactified
to a circle, a type of manifold known as noncommutative cylinder.
The noncommutative cylinder is generally specified by the three
noncommutative coordinates, $\hat t, \hat z $ and ${\hat z}^{\dagger},$ which can be considered as noncommutative versions of
the real parameter $t$ and two complex 
parameters $z = \rho e^{i\phi} $ and its complex conjugate $\bar{ z} = \rho
e^{-i\phi},$  respectively\footnote{In the rest of the paper we assume the notation $x_{\alpha} = (x_0 = t, x_1 = z)
 ~$ and $~ \hat x_{\alpha} = (\hat x_0 = \hat t, \hat x_1 = \hat z), $
 keeping track with the notation in terms of cylinder coordinates.}.
\begin{equation} \label{a1}          
  [\hat z, {\hat z}^{\dagger}]  =  0, \qquad  [\hat t, \hat z]  =  a_0
  \hat z,  \qquad  [\hat t, {\hat z}^{\dagger}]  =  -a_0 {\hat z}^{\dagger}.
\end{equation}
These operators satisfy the same constraint equation as in the
commutative case, namely $\hat z {\hat z}^{\dagger} = {\rho}^2 $.
It is easy to see that the algebra (\ref{a1}) can be recovered from (\ref{1})
by restricting description effectively to only two noncommutative coordinates,
$\hat x_0 \equiv \hat t, \;\; \hat x_1 \equiv \hat z $ and by letting deformation parameter $a_0$
to change into  $-ia_0$.
This simple prescription accommodates for the compactification of the
only space coordinate present in the model.
This algebra (\ref{a1}) admits a class of realizations which has a smooth limit, which is given by 
\begin{eqnarray} \label{a2}
  \hat z  & =  & \rho e^{i\phi} e^{i h(A)}, \\
 {\hat z}^{\dagger}  & =  & \rho e^{-i\phi} e^{-i h(A)},  \\
 \hat t   & =  & t \psi (A) - ia_0 \p_{\phi} \gamma (A),
\end{eqnarray}
with functions $ \psi, h $ satisfying the boundary conditions $h(0) = 0 $ and $ \psi (0) = 1, $ respectively.
Additionally, $h(A)$ is understood to be a hermitian operator
${h(A)}^{\dagger} = h(A).$
Consistency requires that
\begin{equation} \label{a3}
  \gamma (A)  =  \psi (A) \frac{dh}{dA} + 1.
\end{equation}
The shift operator $Z$ is then a unitary operator, $Z^{-1} = Z^{\dagger},$ which is given by
\begin{equation}\label{a4}
Z=e^{+i \int \frac{dt}{\psi(t)}}
\end{equation}
and has the properties
\begin{equation} \label{a5}
  [Z, \hat t]  =  a_0 Z, \qquad  [Z, \hat z]  = [Z, {\hat z}^{\dagger}]  = 0.
\end{equation}
 Note that in Eq. (\ref{a2}), instead of exponential operator $e^{i h(A)},$ one could generally choose 
some function $\varphi (A),$ such that  ${\varphi}^{\dagger} (A) \varphi (A) \ge 0, $
to comply with the general form (\ref{30}).
In this case one would have $\hat z {\hat z}^{\dagger} = {\rho}^2 {\varphi}^{\dagger} (A) \varphi (A). $
There are two particularly interesting situations where the 
 results, together with the implication on statistics, can be given in
 full detail. These situations include choices $i) \; \psi (A) = 1$
 and $ii) \; \gamma (A) = \gamma_0,$ $ \gamma_0$ being some constant. The first choice
 is elaborated for $\kappa$-Minkowski space in \cite{Govindarajan:2008qa},\cite{Meljanac-2},\cite{Meljanac-5} and for the second
 one an appropriate analysis has been done in \cite{Borowiec}. 

\subsection{Star product}

For each $\varphi$-realization and corresponding ordering of the $\kappa$-Minkowski algebra, there exists a unique star product ${\star}_{\varphi},$ twist operator ${\mathcal{F}}_{\varphi}$ and a coproduct ${\Delta}_{\varphi}$.The star product in $\varphi$ realization between two functions $f$ and $g$ in the
algebra of functions on $R^n,$  is generally
given by
\begin{equation} \label{8}
f \; {\star}_{\varphi} \; g ~ = ~  m_{0} ( {\mathcal{F}}_{\varphi} f \otimes g),
\end{equation}
where $ \; m_{0} \; $ is the multiplication map in the Hopf algebra,
 namely, $ \; m_{0} (f \otimes g) = fg$ and ${\mathcal{F}}_{\varphi}$
 is the twist element. 
This can also be written in the form
 \cite{Meljanac-2},\cite{Meljanac-5}
\begin{equation} \label{fstarg}
(f \; \star_{\varphi} \; g)(x)  =   \lim_{\substack{u \rightarrow x }}
 m_0 \left ( e^{x^{\alpha} ( \triangle_\varphi - {\triangle}_{0}) {\partial}_{\alpha} }
    f(u) \otimes g(u) \right ),
\end{equation}
where the coproduct ${\triangle}_{\varphi}$ for translation generators is given in (\ref{63})  and
\begin{equation} \label{untwist}
{\triangle}_{0} (\partial ) = \partial \otimes 1 + 1 \otimes \partial
\end{equation}
is the untwisted coproduct.
The coproduct
 ${\triangle}_{\varphi}$ can also be  obtained by twisting the primitive
coproduct ${\triangle}_{0},$
\begin{equation} \label{cophi}
 {\triangle}_{\varphi} = {{\mathcal{F}}_{\varphi}}^{-1}  {\triangle}_{0}
            {\mathcal{F}}_{\varphi},
\end{equation}
 where the 
twist element ${\mathcal{F}}_{\varphi}$ follows from rearranging
the operator $ e^{x^{\alpha} ({\Delta}_{\varphi} - {\Delta}_{0}) {\partial}_{\alpha}}$
with the help of certain mathematical identity \cite{Govindarajan:2008qa},\cite{Meljanac-5}.

We shall now restrict to the case $i) \; \psi (A) = 1$. The coproduct in this case can be obtained from
Eq. (\ref{63}) by specializing it to $ \psi (A) = 1.$ This coproduct 
then directly determines the star product
in $\kappa$-space, according to relation (\ref{fstarg}), which
finally allows us to identify the corresponding twist operator as
\begin{equation} \label{twist}
 {\mathcal{F}}_{\varphi} = e^{ (N \otimes 1) \ln \frac{\varphi (A \otimes 1 + 1 \otimes A)}{\varphi (A \otimes 1)}
       + (1 \otimes N) (A \otimes 1 +  \ln \frac{\varphi (A \otimes 1 + 1 \otimes
 A)}{\varphi (1 \otimes A)})},
\end{equation}
 where
$N = x_i \partial / \partial x_i \equiv x_i \p_i $ is the dilatation generator with
summation going over space indices only   and $ \; A = -a_0 \p_{0}. $
When written in terms of cylinder coordinates, these operators look as
$N = z \p_z = -i \p_{\phi} ~ $ and $ \; A = -a_0 \p_{t}, $ respectively.

We point out that the star product and the twist element depend
explicitly on the choice of the ordering. For $\phi(A)=e^{-cA}$ where $c \in {\mathbb{R}}$, we obtain a simple interpolation between right ordering ($c = 0$) and left ordering ($c = 1$), with the twist operator given by
\begin{equation} \label{interpol}
{\mathcal{F}}_c = e^{-cN \otimes A + (1-c)A \otimes N}.
\end{equation}
 For  $ \; c = \frac{1}{2}, \; $ we have a symmetric ordering, which is completely different from the totally symmetric Weyl ordering \cite{fedele}.
Using  $\triangle N = N \otimes 1 + 1 \otimes N$, $\triangle A = A \otimes 1 + 1 \otimes A$, and $[N,A]=0$, it is easy to verify that the above class of twist operators ${\mathcal{F}}_c$ satisfies the cocycle condition
\begin{equation}
({\mathcal{F}}_c \otimes 1)(\triangle \otimes 1){\mathcal{F}}_c = 
(1 \otimes {\mathcal{F}}_c)(1 \otimes \triangle){\mathcal{F}}_c,
\end{equation}
for all $c \in {\mathbb{R}}$.

\section{Noncommutative black holes and particle statistics}

Quantum gravity admits the possibility of unusual particle statistics, with associated implications for the spin-statistics connection \cite{sorkin,an1,an2}. Within the paradigm of noncommutative geometry as a possible description of space-time at the Planck scale \cite{dop1,dop2}, and the $\kappa$-Minkowski algebra as a possible description of a class of noncommutative black holes \cite{brian1,brian2,schupp,ohl}, it is interesting to ask how the particle statistics would be described in this framework. 

Let us first consider the commutative flip operator ${\tau}_{0} $ associated with the exchange of particles, given by
\begin{equation} \label{10}
 {\tau}_{0}(f \otimes g) = g \otimes f,
\end{equation}
which satisfies the idempotency property,
 ${{\tau}_{0}}^2 = 1 \otimes 1. $
In the commutative case, the statistics is superselected. This is expressed by the condition that coproduct ${\Delta}_{0} (\Lambda)$ of any generator $\Lambda$ of the Poincare algebra commutes with the flip operator ${\tau}_{0} $, 
\begin{equation}
[{\Delta}_{0} (\Lambda), {\tau}_{0}] = 0.
\end{equation}
In the noncommutative case, the coproduct of  $\Lambda$ is given by
\begin{equation} \label{tcop}
 {\Delta}_{\varphi}(\Lambda) = {{\mathcal{F}}_{\varphi}}^{-1}  {\Delta}_{0} (\Lambda) {\mathcal{F}}_{\varphi}.
\end{equation}
However, the commutative flip operator ${\tau}_{0} $ does not commute with ${\Delta}_{\varphi}(\Lambda)$. If we assume that particle statistics continues to be superselected in the noncommutative case, then we must find a new twisted flip operator which will commute with ${\Delta}_{\varphi}(\Lambda)$. 
Such a twisted flip operator can be defined and is given by
\begin{equation} \label{tflip}
 \tau_\varphi = {\mathcal{F}}_\varphi^{-1} \tau_0 {\mathcal{F}}_\varphi,
\end{equation}
so that  the action of deformed symmetry group respects new statistics
\begin{equation} \label{cond}
[\Delta_\varphi (\Lambda), \tau_\varphi ] = 0.
\end{equation}
Having defined the twisted flip operator, the symmetrization and antisymmetrization can now be carried out with the projection operators $\frac{1}{2} (1 \pm \tau_\varphi )$, respectively. This would ensure that the twisted statistics of a two particle state in the $\kappa$ space would remain unchanged under the action of the twisted symmetry group.

For this particular special class of $\varphi$-realizations, the twist element is given in (\ref{twist}). Using (\ref{twist}) and (\ref{tflip}), we obtain an explicit expression for the twisted flip operator $\tau_\varphi$ for the class of realizations 
$ \psi (A) = 1$ as
\begin{equation} \label{tflip1}
\tau_\varphi % = e^{i(x_iP_i \otimes A - A \otimes x_iP_i)}\tau_0
    = e^{N \otimes A - A \otimes N}\tau_0.
\end{equation}
The last relation makes possible to identify the $R$-matrix and the
corresponding $r$-matrix that is given by
\begin{equation} \label{tflip2}
r  % = \frac{1}{a}  i(x_iP_i \otimes A - A \otimes x_iP_i)
    = \frac{i}{a_0}  (N \otimes A - A \otimes N)
\end{equation}
One can check that this $r$-matrix satisfies the classical Yang-Baxter equation.

%%%%%%%%%%%%%%%%%%%%%%%%

\section{Twisted oscillator algebra in the noncommutative black hole background}

In the commutative theory, probing a black hole geometry with a scalar field provides important information about the space-time structure. Here we initiate a similar study for the noncommutative black hole. In order to study the behaviour of a quantum field around a noncommutative black hole, it is essential to know the algebra of the creation and annihilation operators. Below we obtain the oscillator algebra in the background of a noncommutative black hole, for the particular class of realizations discussed in this Letter.

As a starting point we take the relation
\begin{equation}
f\otimes g=\tau_\varphi (f\otimes g)\label{twistedboson},
\end{equation}
which is consistent with the idempotency property of the statistics
flip operator, ${\tau_\varphi}^2 = 1 \otimes 1.  $
The statistics flip operator ({\ref{tflip1}}) in conjuction with
Eqn.(\ref{twistedboson}) leads to the following condition:
\begin{equation} \label{twistcom}
\phi(x)\otimes\phi(y)-e^{-(A\otimes N-N\otimes A)}\phi(y)\otimes\phi(x)=0.
\end{equation}
It is obvious that condition (\ref{twistcom}) is independent of the
realization.

In order to acquire further insight of the physics of
$\kappa$-cylinder, we stick to the particular choice
$\varphi=e^{-\frac{A}{2}}=e^{-\frac{a_0 \partial_0}{2}}$,
i.e. $h(A) =-\frac{A}{2i}  $. This particular realization corresponds
to symmetric ordering of coordinates and it gives rise to a generalized
Klein-Gordon equation \cite{Govindarajan:2009wt},
\begin{equation}
\left[\partial_{z} \partial_{\bar{ z}} -\frac{4}{a_0^2} {\sinh}^2(\frac{a_0 \partial_0}{2})-m^2\right]\Phi=0,\label{dkg}
\end{equation}
describing a matter field $\Phi (x)$ which probes a black hole background.
The field $\Phi(x)$
%,appearing in the generalized Klein-Gordon equation (\ref{dkg})
 can be decomposed in positive and negative
frequency modes as
\begin{equation} \label{modeexpansion}
\Phi(x)=\int \frac{dp_z}{\sqrt{p_{z}^2+m^2}} \left[b(\omega, {p_z})
 e^{-ip\cdot x}+b^\dagger(\omega, {p_z})e^{ip\cdot x}\right],
\end{equation}
where $b^\dagger(\pm\omega,p_z)=b^\dagger(\mp\omega,p_z)$. Here we
have anticipated that momentum $p$ is a two-vector $p = (p_0, p_z), $
with two  components $p_0 $ and $ p_z $ being canonically conjugated
to cylinder coordinates $t$ and $z,$ respectively. % on cylinder%
The generalized  Klein-Gordon equation (\ref{dkg}) fixes the form of the
dispersion relation
\begin{equation}
p_{0}^\pm=\pm \omega =\pm \frac{2i}{a_0}{\sinh}^{-1}(\frac{a_0}{2i}\sqrt{p_{z}^2+m^2}).\label{P0}
\end{equation}
This expression encodes the on-shell condition for the particles and determines the energies of the positive and negative frequency modes in the Fourier expansion (\ref{modeexpansion}).
In a case we have a product of two bosonic fields $\phi(x)$ and $\phi(y),$ both
 having mode expansion (\ref{modeexpansion}),
 a straightforward application of the Eqn.(\ref{twistcom}) leads to a deformed
 oscillator algebra
\begin{widetext}
\begin{eqnarray}
b^\dagger(p_0, {p_z})b(q_0, {q_z})-e^{ia_0(q_0\partial_{p_z} p_z+\partial_{q_z} q_z p_0)} 
b(q_0,{q_z})b^\dagger(p_0,p_z)=-\delta^{(2)}(p-q),\label{modcom1}\\
b^\dagger(p_0,{p_z})b^\dagger(q_0,{q_z})-
e^{ia_0(-q_0\partial_{p_z}p_z+\partial_{q_z}q_z p_0)} b^\dagger(q_0,{q_z})b^\dagger(p_0,{p_z})=0,\label{modcom2}\\
b(p_0,{p_z})b(q_0,{q_z})-e^{ia_0(q_0\partial_{p_z}p_z-\partial_{q_z} q_z p_0)} b(q_0,{q_z})b(p_0,{p_z})=0.\label{modcom30}
\end{eqnarray}
\end{widetext}
where the time components of the momenta, $p_0$ and $q_0,$ are given in Eqn. (\ref{P0}).

The creation and annihilation operators satisfying the above given deformed commutation 
relations are the ones appearing in the mode decomposition of the scalar field satisfying the generalized Klein-Gordon Eqn.(\ref{dkg}). This generalized Klein-Gordon equation is invariant under the  $\kappa$-Poincar\'e algebra.

%%%%%%%%%%%%%%%%%%%%%%%%%%%%%
Since  deformed statistics flip operator is idempotent, we know 
that the statistics is governed by the permutation group and not by
the more general braided group. In this case the equation (\ref{twistedboson})
identifies the eigenspace of $\tau_\varphi$ corresponding to
eigenvalue $+1,$ that is, it describes deformed bosons, i.e. particles
that obey commutation relations (\ref{modcom1}-\ref{modcom30}).
On the other hand, the projector $\frac{1}{2} (1 - \tau_\varphi )$
determines the eigenspace of  $\tau_\varphi$ with eigenvalue $-1,$
thus defining particles with deformed Fermi-Dirac statistics.
As for condition (\ref{twistedboson}), which was shown to lead to the set of commutatation relations
(\ref{modcom1}-\ref{modcom30}) describing deformed bosons,
the (anti)commutatation relations
obeyed by deformed fermions  can be deduced from the condition
\begin{equation}
f\otimes g= - \tau_\varphi (f\otimes g)\label{twistedfermion}.
\end{equation}
The later condition leads to the same type of relations as (\ref{modcom1}-\ref{modcom30}),
except only for  $-$ sign being  replaced by the $+$ sign.
It gives at least a part of the eigenspace for  $\tau_\varphi$ with
eigenvalue $-1,$ thus describing
  deformed fermions.

The dispersion relation 
\begin{equation}
    p_{z}^2+m^2 = \frac{4}{a_0^{2}}{\sinh}^{2}(\frac{a_0 p_0}{2i})
    \equiv {\epsilon}^{2} \label{dispersion}
\end{equation}
is readily obtained from (\ref{P0}). It is valid for both, deformed bosons and fermions. And even more, with
the help of it we can deduce the form of equation governing dynamics of
deformed fermions by
linearizing \cite{Belhadi:2011mj} the Klein-Gordon equation (\ref{dkg}). Indeed, the dispersion (\ref{dispersion})
has the form of standard energy-momentum relation once we make identification
$\epsilon = \frac{2}{a_0}{\sinh}(\frac{a_0 p_0}{2i}). $ This enables
us to write down the Dirac equation in momentum space as
\begin{equation}
  \epsilon u(p_0, p_z ) \equiv  \frac{2}{a_0}{\sinh}(\frac{a_0 p_0}{2i}) u(p_0, p_z )
 =  (\vec{\alpha} \cdot \vec{p} + \beta m ) u(p_0, p_z ),
  \label{diracequation}
\end{equation}
where $\alpha^i = - \gamma^0 \gamma^i, \beta = \gamma^0 $ are Dirac
$\gamma$ matrices and $u(p_0, p_z )$ is a two-component spinor. Note that in our
 $1+1$-dim case $\vec{\alpha} \cdot \vec{p} = {\alpha}^1 p_z $ and 
$\alpha^i$ and $\beta $ reduce to Pauli matrices and $2\times 2 $ identity matrix, respectively.
The equation (\ref{diracequation}) is not linear any more. It is
nonlinear and its coordinate space counterpart is non-local and thus hard to analyse. The nonlocal operator
can however be expanded in terms of deformation parameter $a_0$ and correspondingly
obtained equation can be analysed within the same order.
This analysis is out of scope of the current paper and is planned for
future publication. Nevertheless, we see that  dynamics of particles in
the background of the noncommutative BTZ black hole is described in terms
of equations that are nonlocal. This happened to be the case for
deformed bosons (see Eq.(\ref{dkg})) as well as for deformed fermions 
 (see Eq.(\ref{diracequation})). The extraction of  any further
information concerning the particle dynamics will thus innevitably require
perturbation calculus.
%%%%%%%%%%%%%%%%%%%%%%%%%%%%%%%%%%%%%%%

The situation described also means  that other objects in our
approach, such as the twist, the $R$-matrix
and the statistics flip operator will eventually have to be deduced through 
the perturbation procedure as well. To illustrate how this can be done,
we investigate a quasi-triangular structure of the symmetry algebra in the
 particular noncommutative background described by
\begin{equation} \label{bc1}
 {\hat x}_i = x_i \bigg( 1 + i\alpha (a\partial)  \bigg) + i\beta (ax){\partial}_i,
\end{equation}
\begin{equation} \label{bc2}
 {\hat x}_0 = x_0 \bigg( 1 + i\delta (a\partial)  \bigg) + ia_0 \gamma
 (x\partial ).
\end{equation}
The above realizations are written perturbatively up to first order in
deformation $a$ and already include the
previously analyzed realizations (\ref{29}) and (\ref{30}) up to the same order. 
Indeed, the later is obtained from the former for $\beta = \gamma + \delta = 0 $.
The quasi-triangular structure is of importance since it gives rise
 to statistical properties of particles in this noncommutative
 background.
To obtain it, we have to find the twist operator, which can be constructed
  once the coproducts for derivatives are known.
The equations (\ref{bc1}),(\ref{bc2}) lead to coproducts
\begin{equation} \label{bc3}
 \Delta ( {\partial}_i) = {\partial}_i \otimes 1 + 1 \otimes {\partial}_i
 - i\gamma a_0 {\partial}_0 \otimes {\partial}_i - i\alpha a_0
 {\partial}_i \otimes {\partial}_0 + {\mathcal{O}} (a^2).
\end{equation}
\begin{equation} \label{bc4}
 \Delta ({\partial}_0) = {\partial}_0 \otimes 1 + 1 \otimes {\partial}_0
 - i(\gamma + \delta ) a_0 {\partial}_0 \otimes {\partial}_0 + i\beta a_0
 {\partial}_i \otimes {\partial}_i + {\mathcal{O}} (a^2).
\end{equation}
The twist element can be now deduced from the exponential factor in (\ref{fstarg}).
However, since there is an ambiguity as where to place the coordinate
variable $x^{\alpha}$ (whether to attach it to the first or to the second 
  factor in the tensor product expansion of the exponential),
 we write the twist element in a sufficiently general form to accommodate 
for this indeterminacy,
\begin{eqnarray} \label{bc5}
 {\mathcal{F}} & =& 1 \otimes 1 - {\lambda}_1 i \beta a_0 x_0
 {\partial}_i \otimes {\partial}_i - (1 - {\lambda}_1) i \beta a_0
 {\partial}_i \otimes x_0 {\partial}_i   \nonumber \\
 & +& {\lambda}_2 i (\gamma + \delta ) a_0 x_0
 {\partial}_0 \otimes {\partial}_0  
 + (1 - {\lambda}_2) i (\gamma + \delta )  a_0
 {\partial}_0 \otimes x_0 {\partial}_0  \nonumber \\
   & -&  {\lambda}_3 i \gamma a_0 x_i
 {\partial}_0 \otimes {\partial}_i - (1 - {\lambda}_3) i \gamma a_0
 {\partial}_0 \otimes x_i {\partial}_i  \\
 & -& {\lambda}_4 i \alpha a_0 x_i
 {\partial}_i \otimes {\partial}_0 - (1 - {\lambda}_4) i \alpha a_0
 {\partial}_i \otimes x_i {\partial}_0 + {\mathcal{O}} (a^2), \nonumber
\end{eqnarray}
where ${\lambda}_1, {\lambda}_2, {\lambda}_3, {\lambda}_4, $ are free parameters,
whose presence reflects the above described indeterminacy. They indicate the weight
with which coordinate $x^{\alpha}$  is attached to the first, i.e. to the second part of the
tensor product in (\ref{bc5}).
 It is now straightforward to
show that coproducts (\ref{bc3}) and (\ref{bc4}) can be reproduced
from primitive coproducts (\ref{untwist}) via twist (\ref{bc5}),
\begin{equation} \label{bc6}
  \Delta ( {\partial}_{\mu}) = {\mathcal{F}}^{-1} ~ {\Delta}_0
  ({\partial}_{\mu}) ~ {\mathcal{F}},
\end{equation}
for any choice of free parameters ${\lambda}_1, {\lambda}_2, {\lambda}_3, {\lambda}_4 $.

The statistics flip operator in this setting can be calculated from
(\ref{tflip}) by using twist element (\ref{bc5}), leading finally  to
the $R$-matrix in the first order in $a_0$, given by
\begin{widetext}
\begin{eqnarray} \label{bc7}
 R & =& 1 \otimes 1  + (2 {\lambda}_1 -1) i \beta a_0 
 \bigg( M_{0i} \otimes {\partial}_i - 
  {\partial}_i \otimes M_{0i} \bigg)
   + \bigg( (2 {\lambda}_1 - 1) \beta + {\lambda}_3 \gamma +
 {\lambda}_4 \alpha - \alpha \bigg) ia_0
 \bigg( x_i {\partial}_0 \otimes {\partial}_i - {\partial}_i \otimes
 x_i {\partial}_0 \bigg) \nonumber \\
& +&   
   \bigg( {\lambda}_4 \alpha + {\lambda}_3 \gamma - \gamma \bigg) ia_0
 \bigg( x_i {\partial}_i \otimes {\partial}_0 - {\partial}_0 \otimes
 x_i {\partial}_i \bigg)
 + (1 - 2 {\lambda}_2) i(\gamma + \delta )a_0
  \bigg( x_0 {\partial}_0 \otimes {\partial}_0 - {\partial}_0 \otimes
 x_0 {\partial}_0 \bigg)
 + {\mathcal{O}} (a^2),
\end{eqnarray}
\end{widetext}
where $M_{0i} = x_0 \p_i - x_i \p_0 $ in the lowest order in $a_0$.
It satisfies quantum Yang-Baxter equation in first order of deformation $a_0,$ for every choice of free parameters
${\lambda}_1, {\lambda}_2, {\lambda}_3, {\lambda}_4 $. Here we can use the known result (\ref{tflip2}) to fix, at least partially, the values of the free parameters.
Thus, by noting that the realization (\ref{29}), (\ref{30}), expanded
to the first order in $a_0$, emerges from the realization (\ref{bc1}),(\ref{bc2})
 under the condition $\beta = \gamma + \delta = 0, $ it is clear that the 
quasi-triangular structure (\ref{bc7}) reduces to 
classical $r$-matrix (\ref{tflip2}) under the same condition, $\beta = \gamma + \delta = 0. $
From this, we can fix the parameters ${\lambda}_3 $ and $ {\lambda}_4, $
${\lambda}_3 = 0, ~ {\lambda}_4 = 1, $ while 
${\lambda}_1 $ and $ {\lambda}_2 $ still remain undetermined.
%What little we
%know of the properties of the twist, we can take ${\lambda}_1 %{\lambda}_2 = 0, $ as these values at this level seem to suit as well as any other.
The analysis thus shows that for a given realization we get
the whole family of $R$-matrices. Are these $R$-matrices  mutualy
equivalent or at least, is there any relation between them is still
an open question that is the subject of our current investigation. We
hope to address these issues in a near future.

Applying the above results to the
$\kappa$-cylinder case, in the first order of deformation parameter
$a_0$ we have the following realizations
\begin{equation} \label{b1}
  {\hat z} = z (1 - \alpha a_0 {\p}_t ) - \beta (a_0 t ) {\p}_z,
\end{equation}
\begin{equation} \label{b2}
 {\hat t} \equiv {\hat x}_0 = t \bigg(1 - (\gamma + \delta) a_0 {\p}_t  \bigg) -i a_0 \gamma {\p}_{\phi},  
\end{equation}
leading to coproducts
\begin{eqnarray}\label{b3}
\Delta (\p_t) & =&  {\Delta}_0 (\p_t) - (\gamma + \delta )
a_0  \p_t \otimes \p_t + \beta a_0  \p_z \otimes \p_z
+ {\mathcal{O}} (a^2), \nonumber \\
\Delta (\p_z) & =& {\Delta}_0 (\p_z) - \gamma a_0  \p_t
\otimes \p_z - \alpha a_0  \p_z \otimes \p_t + {\mathcal{O}} (a^2). \nonumber
\end{eqnarray}
The variable $\phi$ is a polar angle used to parametrize the
cylindrical coordinate $z$.
The relevant $R$-matrix now appears to be
%\begin{eqnarray} \label{b4}
%  R & =& 1 \otimes 1 + \beta a_0  \bigg ( M_{0z} \otimes \p_z - \p_z
%  \otimes M_{0z} \bigg ) +  (\gamma + \beta ) a_0 \bigg ( z \p_0 \otimes \p_z - \p_z
%  \otimes z \p_0  \bigg )   \nonumber \\
% & +& (\gamma + \delta ) \bigg ( a_0 \p_0 \otimes x_{0} \p_0 - x_{0} \p_0
%  \otimes a_0 \p_0  \bigg )
% + \alpha \bigg ( z \p_z \otimes a_0 \p_0 - a_0 \p_0
%  \otimes z \p_z  \bigg ) + {\mathcal{O}} (a^2),
%\end{eqnarray}
\begin{eqnarray} \label{bc8}
 R & =& 1 \otimes 1  + (2\lambda_1 - 1) \beta a_0 
 \bigg( M_{tz} \otimes {\partial}_z - 
  {\partial}_z \otimes M_{tz} \bigg) \nonumber \\
  & +& (2\lambda_1 - 1) \beta  a_0
 \bigg( z {\partial}_t \otimes {\partial}_z - {\partial}_z \otimes
 z {\partial}_t \bigg)  \nonumber \\
& -&   
  i ( \alpha  - \gamma ) a_0
 \bigg(  {\partial}_{\phi} \otimes {\partial}_t - {\partial}_t \otimes
  {\partial}_{\phi} \bigg) \nonumber \\
 & +& (1 - 2\lambda_2) (\gamma + \delta  )a_0
  \bigg( t {\partial}_t \otimes {\partial}_t - {\partial}_t \otimes
 t {\partial}_t \bigg)
 + {\mathcal{O}} (a^2), \nonumber
\end{eqnarray}
where in the lowest order the generators $M_{zt}$ are given as
$M_{zt} = z \p_t - t \p_z $.  It
satisfies Yang-Baxter equation and gives rise to deformed statistics via
\begin{equation} \label{b5}
\phi(x)\otimes\phi(y)- R \phi(y)\otimes\phi(x)=0.
\end{equation}

\bigskip

%%%%%%%%%%%%%%%%%%%%%%%%%

\section{Conclusions}
In this paper we have addressed certain important physics questions that are relevant at the Planck scale. The main physics points discussed here include the nature of particle statistics and the spin-statistics relation at the Planck scale and the deformation of the creation-annihilation operator algebra that lies at the core of any quantum field theory at that scale. 

The issue of particle statistics and the associated spin-statistics connection is an interesting and open problem in quantum gravity. The arguments presented here indicate that around a noncommutative black hole, the particle statistics is twisted and the oscillators describing a free scalar field satisfy a twisted algebra. The twisted oscillator algebra has a very different structure from its commutative counterpart, although it smoothly reduces to the latter when the deformation parameter is taken to zero.

The deformed oscillator algebra around a noncommutative black hole immediately suggests that there could be a breakdown of the spin-statistics relation in this scenario. Such a violation of the spin-statistics relation in quantum gravity has been previously discussed in the context of quantum geons \cite{sorkin,an1,an2}. The violation of the spin-statistics relation observed here is compatible with the fact that the scalar field theory in the background of the $\kappa$-cylinder algebra is nonlocal. In addition, the $\kappa$-Minkowski algebra violates CPT symmetry \cite{Govindarajan:2008qa,Govindarajan:2009wt}. It is thus not totally surprising that the usual spin-statistics relation fails to hold in the description of the Planck scale physics presented here. 
We thus came to the conclusion that the particles in the vicinity of the noncommutative BTZ black hole obey
neither Bose, nor Fermi statistics. However,
for these particles we managed to write down the dynamical equations which govern
their motion in the corresponding background. These equations turn out
to be the deformed Klein-Gordon equation (\ref{dkg}) describing
deformed bosons and the deformed Dirac equation (\ref{diracequation})
describing deformed fermions.

In this Letter we have worked with a specific realization of the
 $\kappa$-Minkowski cylinder algebra, which has a smooth limit as the
 deformation parameter goes to zero. This is an important difference
 compared to the previous treatment of scalar fields  around a
 noncommutative cylinder in \cite{C1,C2}. The twisted flip operator as
 well as the twisted oscillator algebra

Even though we are dealing with black holes at the Planck scale, the K-G equation (41) appears to be relevant for flat space, which may look like a contradiction. However, let us recall that following the work Doplicher et al \cite{dop1,dop2}, the NC setup results from a combined effect of general relativity and quantum uncertainty principle. It is therefore plausible that certain effects of the gravity are already contained in the NC algebras given by eqns. (3), (7) and (8) of this paper. The KG eqn. (41) is a natural consequence of these algebras which is what we have analyzed here. It would however be interesting to consider the NC metric where the ordinary pointwise products are replaced with the star product that has been obtained here, and to consider the scalar field equation in that background. That requires much further analysis of the underlying noncommutative differential geometry including the behaviour of differential forms under the star products obtained in this paper, which is currently under investigation.

It would also be interesting to study physical processes such as the Hawking radiation around a noncommutative black hole, using the oscillator algebra for the scalar fields presented here, which is a work for the future.

{\it{Acknowledgment}}.
     This work was supported by the Ministry of Science and Technology
    of the Republic of Croatia under contract No. 098-0000000-2865.

%%%%%%%%%%%%%%%%%%%%%Appendix%%%%%%%%%%%%%%%%%%%%%%

%\setcounter{equation}{0}
%\setcounter{section}{0}
%\renewcommand{\theequation}{A. \arabic{equation}}
%ovaj \renewcommand{\thesection}{Appendix}
%ovaj \section{Quasi-triangular structure}
%ovaj \vskip 5mm
%\setcounter{section}{0}
%\renewcommand{\thesection}{A}

%%%%%%%%%%%%%%%%

\end{document}